\documentclass[aps,prl]{revtex4-1}

\usepackage{amsmath}
\usepackage{amssymb}
\usepackage{graphicx}
\usepackage[german, english]{babel}

\graphicspath{{Figures/}}

\begin{document}
	
	\title{Physics of the Kitaev model: fractionalization, dynamical correlations, and material connections}

	\author{M. Hermanns$^1$, I. Kimchi$^2$, J. Knolle$^3$}
	\affiliation{$^1$Institute for Theoretical Physics, University of Cologne, 50937 Cologne, Germany}
	\affiliation{$^2$Department of Physics, Massachusetts Institute of Technology, Cambridge, MA, 02139, USA}
	\affiliation{$^3$T.C.M. group, Cavendish Laboratory, J. J. Thomson Avenue, Cambridge, CB3 0HE, United Kingdom}

\begin{abstract}
	Quantum spin liquids have fascinated condensed matter physicists for decades because of their unusual properties such as spin fractionalization and long-range entanglement. Unlike conventional symmetry breaking the topological order underlying quantum spin liquids is hard to detect experimentally. Even theoretical models are scarce for which the ground state is  established to be a quantum spin liquid. The Kitaev honeycomb model and its generalizations to other tri-coordinated lattices are chief counterexamples --- they are exactly solvable, harbor a variety of quantum spin liquid phases, and are also relevant for certain transition metal compounds including the polymorphs of (Na,Li)$_2$IrO$_3$ Iridates and RuCl$_3$. In this review, we give an overview of the rich physics of the Kitaev model, including 2D and 3D fractionalization as well as dynamical correlations and behavior at finite temperatures. We discuss the different materials, and argue how the Kitaev model physics can be relevant even though most materials show magnetic ordering at low temperatures. 
\end{abstract}
	
	\maketitle

\tableofcontents
\section{Introduction}
Quantum spin liquids (QSLs) are among the most enigmatic quantum phases of matter \cite{Anderson,Mossner2001resonating,WenBook,Lacroix2011introduction,Lee2008end,Balents2010spin}. In these insulating magnetic systems, the spins fluctuate strongly even at zero temperature. No magnetic order develops, but the ground state is still far from trivial. The ground state exhibits long-range entanglement \cite{WenBook,TopologicalOrderReview} -- a feature that is often used to identify QSLs theoretically \cite{Jiang2012identifying}. 

Among QSLs, a sub-class often referred to as Kitaev QSLs has recently attracted much attention, both theoretically and experimentally. Indeed, the Kitaev honeycomb model is arguably the paradigmatic example of QSLs because of its unique combination of being experimentally relevant, exactly solvable and hosting a variety of different interesting gapped and gapless QSL phases, not the least a  chiral QSL that harbors nonabelian Ising anyons \cite{Kitaev2006anyons}. 

While the Kitaev interaction was initially believed to be rather artificial, Khaliullin and Jackeli \cite{Khaliullin2005orbital,Jackeli2009} soon realized that it may be the dominant spin interaction in certain transition metal compounds with strong spin orbit coupling, chief among them certain Iridates. To date, several materials have been synthesized that are believed to exhibit Kitaev interactions \cite{Okamoto2007spin,Singh2010antiferromagnetic,Singh2012relevance,Modic2014realization,Plumb2014rucl3,Takayama2015hyperhoneycomb}. Interestingly, the effect may also occur in organic materials \cite{Yamada2016designing} or cold atomic gases \cite{Duan2003controlling}.

Most of the synthesized materials do order magnetically at sufficiently low temperatures
 \cite{Liu2011long,Singh2012relevance,Ye2012direct,Choi2012spin,Biffin2014unconventional,Biffin2014noncoplanar,Sears2015magnetic,Williams2016incommensurate}
--- indicating that while Kitaev interactions are indeed strong \cite{Chun2015direct}, they are not sufficiently strong to stabilize the QSL phase. There are attempts to drive the systems into a QSL phase by applying pressure or by changing the material composition \cite{Takayama2015hyperhoneycomb,Breznay2017resonant}. In addition, if the materials are close enough to the QSL regime, one may hope to find remnants of QSL behavior or  related features from spin fractionalization \cite{Banerjee2016proximate,Banerjee2016neutron,Sandilands2016spin,Nasu2016fermionic}. 
These may appear either at intermediate energies even when the low-energy behavior is determined by the magnetic order, or upon doping mobile charges into the insulator that may then exhibit unusual properties associated with proximate fractionalization \cite{You2012doping,Hyart2012competition,Okamoto2013global,Halasz2014doping}.

In this review, we  give an overview on current theoretical efforts to determine the behavior of Kitaev-based models, not just for the idealized Kitaev interaction and its Kitaev QSL phase but also of the experimentally relevant regimes, to identify experimentally accessible signatures of Kitaev QSLs, and to understand the non-trivial magnetic orders emerging at low temperatures in the various materials. 

This review is structured as follows. In section~\ref{sec:KSL} we discuss the properties of the pure Kitaev model, how to solve it, and what types of $\mathbb Z_2$ QSLs  can occur. We also discuss the finite temperature behavior. In section~\ref{sec:Symmetry} we briefly explain the symmetry properties of materials, and how Kitaev interactions arise. Section~\ref{sec:Dynamical} is concerned with dynamical correlations of Kitaev QSLs, and section~\ref{sec:materials} gives an overview of the relevant materials. We end this review by pointing out some promising directions for  future research.

\section{Kitaev quantum spin liquids}
\label{sec:KSL}
\begin{figure}
	\includegraphics[width=\linewidth]{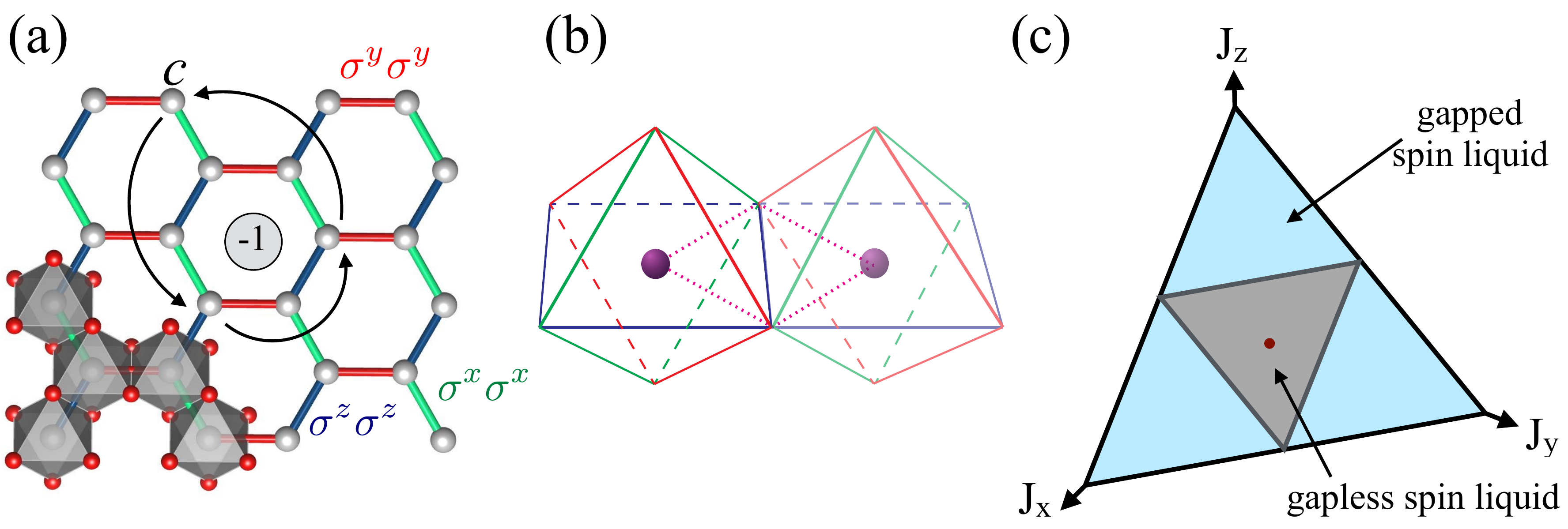}
	\caption{(Color online) (a) Kitaev interactions on the honeycomb lattice. The edge-sharing oxygen octahedra are indicated on the left.  A Majorana fermion encircling a flux $\hat W_p=-1$ gains a $(-1)$ sign to its wave function.  (b) Sketch of the double-exchange path between two neighboring magnetic sites. The green/red/blue planes are \emph{perpendicular} to the $x-$/$y-$/$z-$magnetic axis. (c) Generic phase diagram. Both the nature of the gapless phase in the middle and the precise position of the phase boundaries are \emph{lattice dependent}.  }
	\label{fig:KitaevHoneycomb}
\end{figure}

\subsection{The Kitaev model}
The Kitaev honeycomb model is arguably one of the most important examples of a $\mathbb{Z}_2$ QSL \cite{Kitaev2006anyons}. 
It was originally formulated as spin-$1/2$ degrees of freedom sitting on the vertices of a honeycomb lattice, but it is exactly solvable on any tri-coordinated lattice, independent of (lattice) geometry and spatial dimension \cite{Yao2007exact,Yang2007mosaic,Si2008anyonic,Mandal2009exactly,Kamfor2010kitaev,Hermanns2014quantum,Hermanns2015weyl,Hermanns2015spin-peierls,Obrien2016classification,Rachel2016landau}. 
Nearest neighbor spin degrees of freedom interact via a strongly anisotropic nearest-neighbor Ising exchange \cite{Kugel1982jahn}, where the easy-axis depends on the bond direction as shown in Fig.~\ref{fig:KitaevHoneycomb}(a):
\begin{align}\label{eq:KitaevH}
\hat H&=-\sum_{\langle j,k\rangle} J^{\gamma} K^{\gamma}_{j,k}, 
\end{align}
with the bond operator $K^{\gamma}_{j,k} = \sigma_j^\gamma\sigma_k^\gamma$ if the bond $\langle j,k\rangle$ is of $\gamma$-type. 
The Kitaev interactions along neighboring bonds cannot be satisfied simultaneously, giving rise to `exchange frustration' and driving the system into a QSL phase.\footnote{The \emph{classical} Kitaev model also harbors a (classical) spin liquid \cite{Chandra2010classical,Sela2014order}, which can be described as a Coulomb gas phase \cite{Henley2010coulomb}.  }
Depending on the underlying lattice and the spatial dimension, the Kitaev model \eqref{eq:KitaevH} hosts a variety of both gapped and gapless QSL phases. 
When one of the coupling constants $J^\gamma$ is much larger than the others, the system is in a gapped QSL phase. 
Around the isotropic point $J^x=J^y=J^z\equiv J_K$, however, most lattices harbor an extended gapless QSL, see Fig.~\ref{fig:KitaevHoneycomb}(c).
What types of gapless QSL occur around the isotropic point, and the precise position of the phase transition lines to the gapped phases, are determined by 'projective symmetries' \cite{Wen2002quantum}, see section \ref{sec:projective} below. 
We first give a short overview of how to solve the Kitaev model.
We refer to the original article \cite{Kitaev2006anyons} or the lecture notes by Kitaev and Laumann \cite{Kitaev2009topological} for further details. A detailed discussion on the projective symmetry classification for three-dimensional Kitaev model can be found in Ref.~\cite{Obrien2016classification}.

For each plaquette (i.e. closed loop) in the system, see e.g. the honeycomb plaquettes in Fig.~\ref{fig:KitaevHoneycomb}(a), we can define a \emph{plaquette operator}  
\begin{align}\label{eq:plaquetteOp}
\hat W_p&=\prod_{j\in p} K^\gamma_{j,j+1} .
\end{align}
For a bipartite lattice, where all plaquettes contain an even number of bonds, its eigenvalues are $\pm 1$,  which we refer to as zero (+1) or $\pi$ (-1) flux. 
It is straightforward to verify that all plaquette operators commute with each other and with the Hamiltonian, and thus describe integrals of motion. 
This macroscopic number of conserved quantities allows us to considerably simplify the problem by restricting the discussion to a given flux sector. 
In most of the Kitaev models the flux degrees of freedom are not only static, but also gapped, and we can reduce the discussion to the ground state flux sector. 
Determining which of the exponentially many flux sectors is the one with lowest energy is often non-trivial. 
For lattices with mirror symmetries that do not cut through lattice sites, we can make use of Lieb's theorem \cite{Lieb1994flux}, which states that plaquettes of length 2 mod 4  carry zero flux in the ground state, while plaquettes of length 0 mod 4 carry $\pi$ flux. 
Unfortunately, Lieb's theorem is not applicable for most of the three-dimensional tri-coordinated lattices, and one needs to verify the ground state flux sector numerically. Interestingly,  Lieb's theorem nevertheless gives the correct prediction (with very few exceptions), even though it is strictly speaking not applicable \cite{Obrien2016classification}.

Let us now represent the spin degrees of freedom by four Majorana fermions as
\begin{align}\label{eq:4Maj}
\sigma_j^\alpha=ia_j^\alpha c_j, \mbox{ with } \{a_j^\alpha,a_k^\beta\}=2\delta_{j,k}\delta_{\alpha,\beta}, 
\,\,\{c_j,c_k\}=2\delta_{j,k}, \mbox{ and } \{a_j^\alpha, c_k\}=0,
\end{align}
where $j$ denotes the site index and $\alpha$ the spin component. 
This enlarges the Hilbert space on each site from dimension 2 to 4, but we can recover the physical Hilbert space by requiring that the spin algebra is faithfully reproduced. 
More formally, this is achieved by a projection operator $\mathcal{P}_j=\frac 1 2 (1+a_j^x a_j^y a_j^z c_j^{\phantom x}) $ for each lattice site, which projects generic states to the local physical Hilbert space.
Using this reformulation of the spins, the bond operators are given by $K^\gamma_{j,k}=-(ia_j^\gamma a_k\gamma) i c_j c_k\equiv -i \hat u_{j,k} c_{j} c_{k}$. 
At first glance, this seems not to simplify our discussion, because the Hamiltonian consists now purely of quartic terms. 
However, the bilinear operators $\hat u_{j,k}$ commute with each other as well as with any bilinear operator containing the $c$ Majoranas, and we can replace them by their eigenvalues $\pm 1$. 
This effectively reduces \eqref{eq:KitaevH} to a non-interacting Majorana hopping Hamiltonian in a static background $\mathbb{Z}_2$ gauge field. 
Note that the eigenvalues of the $\hat u$ operators themselves are not physical; only the gauge-invariant plaquette operators $\hat W_p=\prod_{j\in p} (-i \, \hat u_{j,j+1})$ yield physical quantities. 
In fact, the projection operator acting on a site $j$ flips all the $\hat u$ operators emanating from this site. 
Fixing the eigenvalues of $\hat u$ should, therefore, be considered as `fixing a gauge'. 
As long as we compute gauge-invariant quantities, gauge-fixing is (mostly\footnote{See however the discussions in Ref.s~\cite{Pedrocchi2011physical,Zschocke2015physical} on the effects of the projection on physical quantities.}) harmless, and one often does not need to perform the projection to the physical subspace explicitly. 

\subsection{Classifying Kitaev quantum spin liquids by projective symmetries}
\label{sec:projective}

When one of the coupling constants dominates, the Majorana system is gapped and the low-energy degrees of freedom are the flux excitations of Eq.\eqref{eq:plaquetteOp}. The effective Hamiltonian is identical (in 2D) or at least similar (in 3D) to that of the Toric Code \cite{Kitaev2003fault,Hamma2005string}. 
Around the isotropic point, the fluxes are still gapped, but the Majorana system is generically gapless and, thus, determines the low-energy properties of the Kitaev QSL.  

We now restrict the discussion to the ground state flux sector and analyse the properties of the Majorana system. 
In close analogy to electronic systems, Majorana fermions can form a variety of gapless or gapped band structures. In the following, we will call gapless systems (semi-)metallic, even though Majorana fermions are chargeless and there is consequently no $U(1)$ symmetry -- only parity is a good quantum number.
The properties of the Majorana system are determined not by the bare symmetries of the spin system, but by the projective symmetries \cite{Wen2002quantum}. 
Because of the emergent $\mathbb Z_2$ gauge field, the effective Majorana system needs to obey symmetries only up to gauge transformations. 
As a result, each symmetry can be implemented in two distinct ways:  either they are implemented exactly as in electronic systems, or the gauge transformation artificially doubles the unit cell, and thus shifts the symmetry relations in momentum space by half a reciprocal lattice vector. 
The former will be denoted as trivial implementation, the latter as non-trivial.
For instance,  time-reversal always needs to be supplemented with a sub-lattice symmetry in order to be a symmetry of the Majorana system.\footnote{On non-bipartite lattices, the system spontaneously breaks time-reversal symmetry and the ground state will be two-fold degenerate \cite{Kitaev2006anyons,Yao2007exact}.}
Either the sub-lattice symmetry can be implemented identically for each unit cell (such as for the honeycomb lattice \cite{Kitaev2006anyons}) or it needs to be staggered for neighboring unit cells (such as for the square octagon lattice \cite{Yang2007mosaic, Obrien2016classification}). The latter causes a shift in the momentum space by half a reciprocal lattice vector $\mathbf k_0$:
\begin{align}
\hat h(\mathbf k)&=  U_{\mbox{\tiny T}} \, \hat h^\star(-\mathbf k+\mathbf k_0)\, U_{\mbox{\tiny T}}^{-1}&\qquad
\epsilon(\mathbf k)& =  \epsilon(- \mathbf k + \mathbf k_0).
\label{eq:TR}
\end{align}
In 2D systems, $\mathbf k_0=0$ implies that Dirac cones are \emph{stable}\footnote{Here, stable means that one can make an arbitrary small change of the Kitaev couplings without gapping the system.}, but Majorana Fermi lines are not, while for $\mathbf k_0\neq 0$ the situation is reversed: Majorana Fermi lines are stable, but Dirac cones are not. This lies at the heart of the different behaviors of the Kitaev QSLs on the honeycomb \cite{Kitaev2006anyons} and the square-octagon lattice \cite{Yang2007mosaic,Lai2011su2}.

\begin{figure}
	\includegraphics[width=\linewidth]{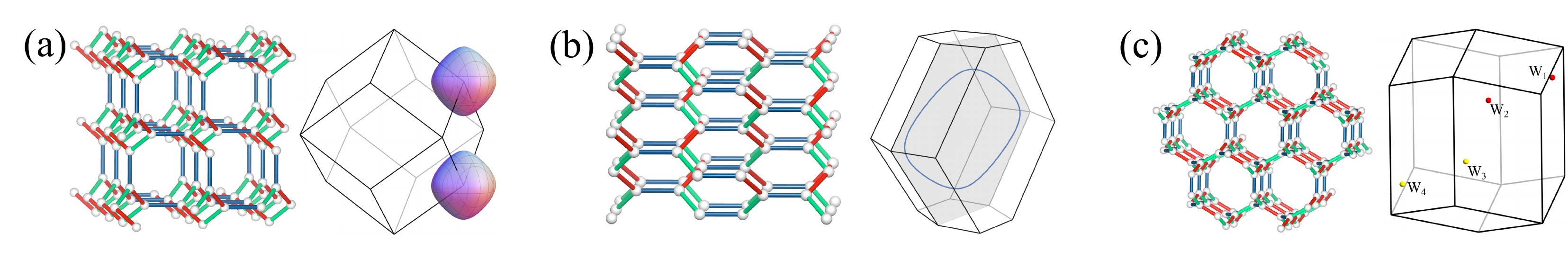}
	\caption{(Color online) Kitaev QSLs with (a) Majorana Fermi surfaces, (b) a nodal line, and (c) Weyl nodes, which are realized for the Kitaev model on the (a) (10,3)a (hyperoctagon), (b) (10,3)b (hyperhoneycomb), and (c) (8,3) b lattice \cite{Obrien2016classification}.  }
	\label{fig:3DKSL}
\end{figure}

Also in 3D, the Kitaev model shows rich physics; depending on the underlying lattice structure, one can realize Kitaev QSLs with any type of band structure ranging from Majorana Fermi surfaces, over nodal lines and Weyl points, to gapped states. Remarkably, the band structures are \emph{generically topological}, i.e. they are characterized by a topological invariant and/or possess topologically protected surface modes \cite{Schaffer2015topological,Obrien2016classification}, in close analogy to electronic systems \cite{Schnyder2008classification,Kitaev2009periodic,ChiuReview}.
If time-reversal symmetry is implemented trivially, the only stable zero modes are  nodal lines (3D), even though there may be additional features, such as symmetry-protected flat bands or Dirac cones at the isotropic point \cite{Obrien2016classification,Yamada2017}. 
If time-reversal symmetry is implemented non-trivially, the QSL generically harbors stable Majorana Fermi surfaces \cite{Hermanns2014quantum,Hermanns2015spin-peierls,Obrien2016classification}. 
An interesting situation arises when time-reversal is implemented non-trivially, but the lattice also has a trivially implemented inversion symmetry. In this case, the only stable zero-energy modes are Weyl nodes \cite{Hermanns2015weyl}. 
Examples of these three different types of spin liquids are shown in Fig.~\ref{fig:3DKSL}.

The projective symmetry analysis does not only determine the physics for the pure Kitaev interaction, but also how the Kitaev QSL responds to perturbations. As the flux excitations are gapped, the Kitaev QSL is stable for a finite range, but its nature may change. For instance, while applying an external magnetic field does not change the qualitative features of Majorana Fermi surfaces and Weyl points, it generically gaps nodal lines into Weyl points, and thus drives the system into a Weyl spin liquid phase \cite{Hermanns2015weyl}. Interactions between Majorana fermions are  irrelevant (in the renormalization group sense) for nodal lines and Weyl points \cite{Lee2014heisenberg}, but partially gap the Majorana Fermi surface to nodal lines \cite{Hermanns2015spin-peierls}.

\subsection{Confinement and finite temperature}
So far, the discussion has been restricted to zero temperature. But the special properties of the Kitaev model allows us to also understand the finite temperature behavior, which is intimately related to the physics of confinement-deconfinement. 
In Kitaev's exact solution, the gauge field is static and the emergent Majorana fermions are deconfined, meaning they can be described as true quasiparticles. Transitions out of the QSL, namely confinement of the Majorana fermions, occur via the flux excitations of the emergent $Z_2$ gauge field. 

The mechanism for confinement can be  seen as follows. Consider the complex quantum amplitude for the process of hopping a Majorana fermion from site $i$ to site $j$, equivalently the matrix element for the transition from occupancy of $i$ to occupancy of $j$. This process entails taking the total sum of the complex amplitudes for all possible paths from $i$ to $j$.
Consider two such paths: 
If there is an odd number of fluxes $\hat W_p=-1$ in the region enclosed by them, then their amplitudes will have a relative $(-1)$ sign, resulting in complete destructive interference.
This is simply the emergent-gauge-field analog of the Aharonov-Bohm effect. 
Confinement transitions out of the spin liquid arise through this nontrivial mutual phase factor between fluxes and emergent fermions:
 (i) At zero temperature, confinement occurs when the Hamiltonian is modified enough so as to condense the fluxes.\footnote{Flux condensation implies that the flux numbers $\hat W_p$ are in a coherent superposition of $+1$ and $-1$ and are no longer good quantum numbers.} This requires a finite perturbation since the fluxes are gapped.  (ii) At finite temperatures, confinement occurs when the fluxes are thermally excited at finite density. 

The $T>0$ confinement transitions out of the QSL are different in two dimensions versus three dimensions. 
In 2D, fluxes are point objects with a gap $\Delta$ that is determined by the underlying lattice and the Kitaev couplings (e.g. $\Delta\approx .26 J_K$ for the honeycomb model at the isotropic point \cite{Kitaev2006anyons,Pachos2007wavefunction}). 
One might imagine that $\Delta$ determines a finite-temperature confinement transition, since for $T<\Delta$, the typical separation between fluxes is exponentially large, and only for $T>\Delta$ do the fluxes proliferate with a high probability on all plaquettes. However, it is known that 2D gauge theories are confining at any nonzero temperature \cite{WenBook}, which here can be understood as the Boltzmann weight giving fluxes an exponentially-small but  finite density. In 2D, the Majorana fermions are confined at any nonzero temperature $T>0$.

In 3D, however, $Z_2$ gauge theories have a deconfined phase that extends to small finite temperatures \cite{Senthil2000Z2gauge}.
Here, the fluxes are no longer point objects with a finite gap, but rather  closed flux loops with an energy that depends on the length of the loop. 
At small temperatures, fluxes are excited, but stay small because of the loop tension. Only for sufficiently large temperatures will the loops become large and span the full system. 
This gives a thermodynamic transition at a temperature $T^\star$ -- determined by the effective loop tension of the flux loops -- which confines the Majorana fermions and drives the system out of the 3D QSL\cite{Nasu2014vaporization,Kamiya2015magnetic,Kimchi2014three}. How do we compute the tension of a flux-line, say at zero temperature in the ground state? Though it may at first seem counterintuitive, the tension of flux lines is given by the energy of the Majorana fermions, hopping in different static configurations of a gauge field.  
Indeed this is a hallmark of fractionalization: the presence of deconfined quasiparticle excitations requires having well-defined excitations of the gauge field, and vice-versa.

The entire spectrum of the pure Kitaev models has also been computed numerically in terms of the fluxes and Majorana fermion variables, which permits a study of thermodynamic quantities via Monte Carlo sampling over the static $Z_2$ flux sectors \cite{Nasu2014vaporization,Kamiya2015magnetic,Nasu2015thermal}. Below $T^*$ -- corresponding to the flux gap $\Delta$ in 2D or the loop tension in 3D as discussed above --  fluxes are (approximately) frozen out and the characteristic properties of the Kitaev QSL emerges, e.g. the linear Dirac density of states of Majorana fermions. 
The numerical simulations also show a larger scale, $T^{**}$, corresponding to bare Kitaev exchange energy $J_K$. 
The paramagnet above the confinement temperature of the spin liquid is adiabatically connected to the high-temperature $T\gg J_K$ paramagnetic phase. However, below $T^{**}\approx J_K$ there is a cross-over into an intermediate correlated paramagnetic regime: the nearest neighbor spin correlations of the Kitaev exchange develop.  This is seen in the specific heat of the isotropic 2D honeycomb model which shows two pronounced crossover peaks at both $T^*$ and $T^{**}$ with a linear in $T$ behavior in between \cite{Nasu2015thermal}. For magnetic phases proximate to the spin liquid, similar $T^{**} \sim J_K$ cross-overs from the uncorrelated to the correlated paramagnet have been seen in numerical studies \cite{Youhei2016clues}. This suggests that qualitative features of the correlated Kitaev paramagnet can survive in currently existing materials.

\section{Symmetry and chemistry of the Kitaev exchange}
\label{sec:Symmetry}

In solid state materials, the Kitaev couplings were originally proposed for 2D systems where magnetic spin-half sites occupy the sites of a honeycomb lattice \cite{Khaliullin2005orbital,Jackeli2009,Chaloupka2010kitaev}, see Fig.~\ref{fig:KitaevHoneycomb}(a).
Importantly, the magnetic superexchange between two adjacent sites has to involve more than one oxygen exchange pathway. The magnetic site, Iridium, is octahedrally coordinated by six oxygen atoms forming the vertices of an octahedron. These octahedra are edge-sharing, so that there are exactly two oxygens between a given pair of Ir sites, with Ir-O-Ir bonds forming a 90 degree angle, see Fig.~\ref{fig:KitaevHoneycomb}(b). 

In this edge-sharing-octahedra geometry, it can be shown that within the single-band Hubbard model associated with the effective S=1/2 manifold, the hopping of electrons between Ir sites is completely forbidden. This occurs due to a complete destructive interference between the two Ir-O-Ir exchange pathways, resulting from the combination of the geometry and spin-orbit-coupling (SOC). The SOC allows an effective magnetic field for an electron with a given spin, permitting imaginary hopping amplitudes, and indeed the two paths have opposite amplitudes of $i$ and $-i$. 

It then becomes necessary to consider exchange involving multiple bands, i.e. higher excited  multiplets, in order to derive a nonzero value for the interactions among the low-energy $S=1/2$ degrees of freedom. Let us consider all terms that are symmetry allowed. Obviously the Heisenberg term will be generated, as well as a ``pseudo-dipole" exchange term, which couples the component of spin lying along the bond between the two sites.
\footnote{In the literature, this pseudo-dipole exchange has been denoted as a bond-Ising exchange ``$I$" when all bonds with this term share the same orientation \cite{Kimchi2015unified}; it has also been re-combined with the Heisenberg and Kitaev exchanges into an equivalent symmetry-allowed term, off-diagonal in the basis of the Kitaev exchange, denoted as ``$\Gamma$" \cite{Rau2014generic}. }
However there is also a third term allowed by symmetry, which is the Kitaev exchange: it couples the component of spin $\gamma$ which is perpendicular to the plane formed by the two exchange paths between the Ir atoms, as depicted in Fig.~\ref{fig:KitaevHoneycomb}(b). 
In certain parameter regimes, the Kitaev exchange may dominate, but the  nearest neighbor Hamiltonian can often be summarized as \cite{Jackeli2009, Chaloupka2010kitaev, Chaloupka2015hidden, Khaliullin2005orbital, Valenti2013ab, Rau2014generic, Mazin2013origin, Winter2016challenges, Kateryna2013ab, Kim2015kitaevmagnetism}
\begin{equation}
H_{i j} = I (\vec{S}_i \cdot \vec{r}_{i j}) (\vec{S}_j \cdot \vec{r}_{i j} )
+ J_H (\vec{S}_i \cdot \vec{S}_{j})
 + J_K (\vec{S}_i \cdot \vec{\gamma}_{i j}) (\vec{S}_j \cdot \vec{\gamma}_{i j} )
\end{equation}
where $\vec{r}_{i j}$ is the unit vector connecting sites $i$ and $j$, and the Kitaev label is $\vec{\gamma}_{i j} \propto \vec{r}_{\text{Ir-O}_1} \times \vec{r}_{\text{Ir-O}_2} $. 
Note that the magnitude of the pseudo-dipole $I$, Heisenberg $J_H$, and Kitaev $ J_K$ coefficients can be different on bonds that are symmetry-distinct. 

The Kitaev interaction can also be generalized to materials with other lattices,  see e.g. \cite{Avella2015quantum,Reuther2012magnetic, Rousochatzakis2016kitaev, Becker2015spinorbit, Lee2014heisenberg, Lee2014order, Lee2015theory, Kimchi2014Kitaev, Kimchi2014three, Kimchi2015unified}. It is then immediately important to note that the Kitaev term is very different from the pseudo-dipole term in two ways. First,  they involve spin exchange in different directions, where the Kitaev axes $x,y,z$ are all orthogonal to each other (Fig.~\ref{fig:KitaevHoneycomb} (b)), in contrast to the various orientations of the bonds.  Second, unlike the pseudo-dipole term whose exchange vector is linearly related to the bond orientation, the Kitaev exchange axis does not have the symmetry transformation properties of a vector: if one bond is related to another bond by some rotation matrix $R$, their Kitaev exchange axes are \emph{not} generally related by the rotation $R$. Rather, the Kitaev exchange transforms under spatial rotations as an $L=2$ tensor form, involving magnetic sites as well as bonds \cite{Khaliullin2001order, Jackeli2009, Kimchi2014Kitaev} (see Fig.~\ref{fig:KitaevHoneycomb}). 
For materials with edge-sharing octahedra, the  spatial orientations of Kitaev exchanges can be determined by considering such lattices as sub-lattices of the FCC lattice formed by a dense octahedral tiling. Prominent examples include the hyperkagome \cite{Okamoto2007spin} lattice of Na$_4$Ir$_3$O$_8$, the hyperhoneycomb and stripyhoneycomb lattices of $\beta$-Li$_2$IrO$_3$ \cite{Takayama2015hyperhoneycomb} and $\gamma$-Li$_2$IrO$_3$ \cite{Modic2014realization} respectively, as well as the layered honeycomb lattices of RuCl$_3$ \cite{Plumb2014rucl3}, Na$_2$IrO$_3$ and $\alpha$-Li$_2$IrO$_3$ \cite{Singh2010antiferromagnetic,Singh2012relevance}.

\section{Dynamical Correlations}
\label{sec:Dynamical}

A typical property of QSLs is the absence of rotational or translational symmetry breaking.  
But clearly, the lack of evidence for long range spin correlations at low temperatures cannot be taken as evidence for its absence, e.g. because of strong quantum fluctuations \cite{Lacroix2011introduction}. 
Another defining feature of QSLs is long-range entanglement and fractionalization of quantum numbers, which for gapped QSLs can be described mathematically as topological order which entails topological ground state degeneracy. These properties can be calculated exactly for the Kitaev model \cite{Yao2010entanglement,Dong2008exact,Lahtinen2009non,Feng2007topological,Baskaran2007exact}. However, these features are not easy to probe directly in experiment. It is therefore useful to also consider non-universal features which can still shed light on the physics, especially when connecting to experiments. In the following, we characterize the exactly soluble point of the Kitaev QSL through its dynamical correlation functions, which are  relevant for inelastic scattering experiments. We discuss the robustness of these features to perturbations within the QSL phase, as well as across phase transitions to nearby orders, and the relation to current experiments.

\subsection{Static Correlations and Selection Rules}
Spin correlations in the Kitaev QSL are short-ranged and vanish exactly beyond nearest neighbors $S^{ab}_{ij} =  \langle\sigma^a_i \sigma^b_j \rangle \propto \delta_{a,b} \delta_{\langle i,j\rangle_a}$.
There is a strong spin anisotropy such that along an $a$-type bond $\langle i,j\rangle_a$ only the $S^{aa}_{ij}$ component is non-zero indicated by the symbol $\delta_{\langle i,j\rangle_a}$. Of course a short decay length of spin correlations is expected for a QSL but the ultra short ranged nature of the Kitaev model is special and directly related to the fact that spins fractionalize into a Majorana fermion and a nearest neighbor pair of gapped static $\pi$-fluxes \cite{Baskaran2007exact} -- the first spin operator creates two fluxes sharing a $\langle i,j\rangle_a$ bond and since flux sectors are orthogonal the second spin needs to remove the very same fluxes for a non-zero matrix element. 
This constraint is removed by additional perturbations in the Hamiltonian.  Whether it leads to exponentially decaying spin correlations (e.g. by a Heisenberg term) or algebraically decaying ones (e.g. by a magnetic field) can be determined from modified selection rules, namely whether a pair of fluxes can be locally neutralized by the perturbation \cite{Tikhonov2011power,Mandal2011confinement}. 

Correlations of operators diagonal in fluxes, e.g. the energy-energy correlator \cite{Lai2011powerlaw} 
\begin{align}
C_{ij} = \left[ \langle \sigma^z_i \sigma^z_{i'} \sigma^z_j \sigma^z_{j'} \rangle - \langle \sigma^z_i \sigma^z_{i'} \rangle \langle \sigma^z_j \sigma^z_{j'} \rangle \right] \delta_{\langle i,i'\rangle_z} \delta_{\langle j,j'\rangle_z}\nonumber,
\end{align}
are only determined by the Majorana sector. It changes its qualitative behavior across the QSL transitions, decaying algebraically in the gapless phases, e.g. $C_{ij} \propto \frac{1}{|\mathbf{r}_i-\mathbf{r}_j|^4}$ from the Dirac spectrum on the honeycomb lattice, but exponentially in the gapped phases \cite{Yang2008fidelity}. 
Remarkably, the qualitative behavior of static correlations in exactly soluble Kitaev models is independent of dimensionality or lattice details - the static nature of the emergent $Z_2$ gauge field entails the same selection rules.

\subsection{Dynamical correlations of the Kitaev spin liquid}

Dynamical correlations are directly related to experimental observables. For example, the  spin structure factor $S^{ab}_{\mathbf{q}}$, which is the Fourier transform in space and time  of the dynamical spin correlation function
$S^{ab}_{ij}(t) =  \langle\sigma^a_i (t) \sigma^b_j (0) \rangle$,
is directly proportional to the cross section of inelastic neutron scattering (INS) experiments. 

The dynamical spin correlation function can be expressed entirely in terms of Majorana fermions \cite{Baskaran2007exact}. The role of the fluxes is incorporated by a sudden perturbation of the Majorana fermions which turns the calculation of the {\it dynamical equilibrium} correlation function into a true {\it non-equilibrium} problem
\begin{align}
S^{ab}_{ij} (t)&=   - i \langle M_0|e^{i H_0 t} c_i e^{-i \left(H_0+V_a\right) t} c_j |M_0 \rangle \delta_{a,b} \delta_{\langle i,j\rangle_a}.
\label{eq:DynamicsS}
\end{align}
Here, $|M_0\rangle$ is the ground state of the Majorana sector in the flux free sector described by $H_0$ and the perturbed Hamiltonian $\left(H_0+V_a\right)$ differs only in the sign of the Majorana hopping on the $a$-bond from the extra pair of fluxes. The problem turns out to be a local quantum quench related the famous X-ray edge problem \cite{Nozieres1969singularities}. It can be evaluated exactly even in the thermodynamic limit \cite{Knolle2014dynamics,Knolle2015dynamics,Knolle2016dynamics}. 

\begin{figure}
	\includegraphics[width=\linewidth]{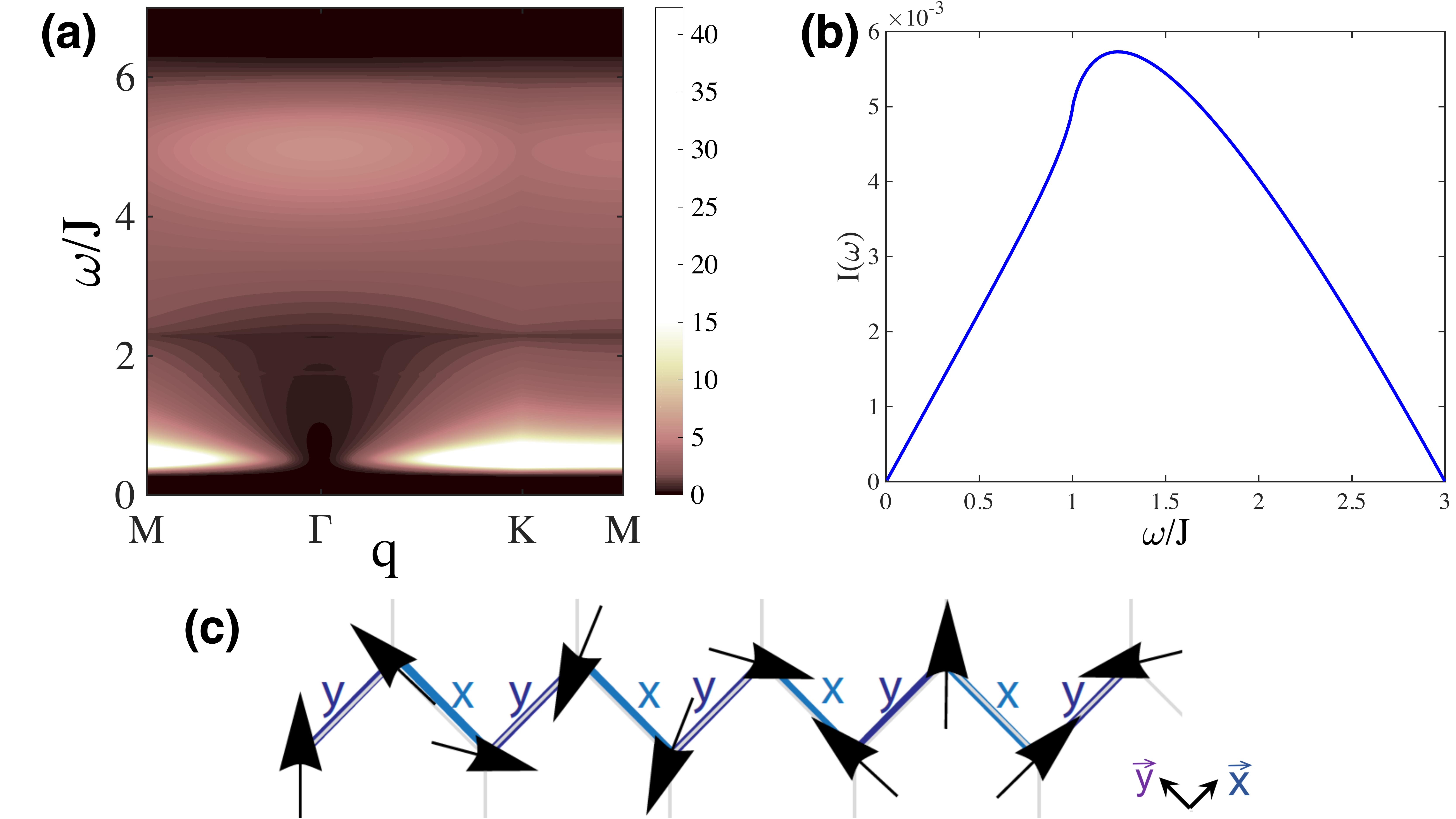}
	\caption{(Color online) The left panel (a) shows the dynamical structure factor $S_{\mathbf{q}}(\omega)$ at zero temperature for the antiferromagnetic isotropic Kitaev honeycomb model along a path in the BZ \cite{Knolle2015dynamics}. Even for the gapless QSL phase the response is gapped and the broad high frequency features are determined by the Majorana fermion DOS. The right panel (b) shows the corresponding Raman intensity which is independent of photon polarization \cite{Knolle2014raman}.
	Panel (c) depicts one possible ordered state that can result from additional interactions beyond  Kitaev exchange, the counterrotating spiral order seen in $\alpha,\beta,\gamma$-Li$_2$IrO$_3$. Its correlations are neither ferromagnetic nor antiferromagnetic: to uncover the ``Kitaev" pattern of correlations \cite{Kimchi2015unified}, tilt your head at $45^\circ$, and observe how $x$-bonds ($y$-bonds) have aligned $S^x$  ($S^y)$ but anti-aligned $S^y$ ($S^x)$. } 
	\label{fig:KitaevStructureFactor}
\end{figure}

The main qualitative features of the spin structure factor are again independent of dimensionality and lattice details. As a concrete example, we show in panel (a) of Fig.\ref{fig:KitaevStructureFactor} $\sum_a S^{aa}_{\mathbf{q}}(\omega)$ of the isotropic AFM honeycomb Kitaev model along a representative path in the Brillouin zone (BZ) \cite{Knolle2015dynamics}. The low energy response has a gap (here $\Delta\approx0.26 J_K$), even in the presence of gapless Majorana fermions, because spin flips always excite gapped fluxes. It is remarkable that INS would be able to directly measure the energy it costs to excite a nearest neighbor flux pair. Above the gap the response is governed by the Majorana DOS. For example in Fig.~\ref{fig:KitaevStructureFactor}(a) suppression of spectral weight just above $\omega=2J_K$ is a direct consequence of a van Hove singularity in the DOS and the sharp drop of intensity above $\omega=6K$ stems from the Majorana bandwidth\cite{Knolle2014dynamics}. Remarkably, the low frequency response is similar on all lattices: if the Majorana DOS vanishes $S_{\mathbf{q} } (\omega)$ follows the same asymptotic power law; if the DOS is constant towards zero energy, e.g. from a Majorana Fermi surface, then  $S_{\mathbf{q} } (\omega) \propto \left(\omega-\Delta \right)^{-\alpha}$ diverges with an X-ray edge exponent $\alpha>0$ \cite{Smith2015neutron,Smith2016majorana}.
 This separation of features from either of the two emergent excitations reveals more direct signatures of Kitaev QSL physics in the structure factor as normally expected for a fractionalized system.  

An alternative probe is magnetic Raman scattering -- inelastic light scattering in the meV range -- probing correlations between two-photon events \cite{Fleury1968scattering}. 
Due to the different selection rules Raman scattering does not excite fluxes but pairs of Majoranas which allows an exact calculation for two- \cite{Knolle2014raman,Perreault2016resonant,Perreault2016raman,Perreault2016majorana} and three-dimensional lattices \cite{Perreault2015theory}. The asymptotic low frequency response $I (\omega)$ is a direct probe of the low energy DOS, for example linear in frequency  for the isotropic honeycomb lattice shown in Fig.\ref{fig:KitaevStructureFactor} (b). 
Yet another probe is resonant inelastic X-ray scattering (RIXS), which is in principle able to probe both types of fractionalized sectors \cite{Halasz2016resonant}. 

What is the effect of small perturbations deviating from the pure Kitaev point but remaining inside the QSL phase? Perturbation theory around the integrable point shows how the selection rules are modified \cite{Tikhonov2011power,Mandal2011confinement}. For example, the flux gap in the structure factor of the gapless Kitaev QSLs is removed by a direct coupling of spin flip processes to pairs of Majoranas \cite{Song2016low}. However, the main features of the response are expected to be robust on general grounds because the Kitaev QSL is a stable phase persistent over a finite range of perturbing interactions \cite{Hong2011possible,Chaloupka2010kitaev, Chaloupka2013zigzag,Kazuya2015density}.
This is due to the gap of the emergent gauge field in conjunction with the vanishing DOS of Majorana fermions, which renders fermion-fermion interactions irrelevant (except on certain three dimensional lattices \cite{Hermanns2015spin-peierls}).

Dynamical properties have also been calculated at nonzero temperature, e.g. the Raman scattering signal \cite{Nasu2016fermionic} or the structure factor \cite{Yoshitake2016fractional}. Both change their qualitative behavior at the characteristic cross-over scales $T^*\sim\Delta$ and $T^{**} \sim J_K$.
Another important deviation from the Kitaev point is the addition of defects. Several works have shown that the response to static disorder can reveal Kitaev QSL features \cite{Willans2010disorder,Willans2011site,Dhochak2010magnetic,Vojta2016kondo,Sreejith2016vacancies,Halasz2016coherent}.

\subsection{Spin dynamics from Kitaev magnetism proximate to the QSL}
Now consider the magnetically ordered phases that are proximate to the Kitaev QSL, i.e.\ consider a Hamiltonian with sufficient non-Kitaev exchanges so that the QSL phase is destroyed and the Majorana fermions become confined.
Recent ED studies \cite{Youhei2016clues} and the time dependent density matrix renormalization group \cite{Gohlke2017dynamics} indicate that the broad high frequency features of the structure factor, as computed for the Kitaev model, are preserved in these proximate phases with long-range ordered magnetism. 
These high-frequency features have also been interpreted in terms of  multi-spin-wave-based excitations above the magnetically ordered phases.  
As elaborated below, there are two main magnetically ordered phases that appear in the Kitaev-type materials: collinear zigzag antiferromagnetic order, and an unusual counterrotating spiral order. The magnon spin dynamics have been computed for both orders
via model Hamiltonians which include Heisenberg exchanges in addition to a strong Kitaev exchange.  The details are different (e.g.\ the spiral entails magnon Umklapp scattering from the  Kitaev exchange), but in both cases, magnon bands show the unusual feature of a high-$\omega$, low-$q$ peak in intensity 
\cite{Choi2012spin,Kimchi2016spin}.
This unusual signal  can be understood intuitively via the Klein duality \cite{Khaliullin2002quantum, Chaloupka2010kitaev,Chaloupka2015hidden,  Kimchi2014Kitaev, Kimchi2016spin} (elaborated below) relating certain Kitaev-based and Heisenberg-based models, which maps wavevector $q$ to  $\pi-q$ in appropriate units.  
The conventional Heisenberg magnon spectrum, with large intensity at high frequency for $q$ near the BZ boundary, is then flipped across $q$ space to produce the intensity at high frequencies near the zone center $\Gamma$ point.
Magnon breakdown and multi-magnon processes are also expected to arise in these materials \cite{Winter2017breakdown}. This can be seen via a strong coupling of one-magnon and two-magnon states, which was shown to lead to a broad band of intensity centered at a high frequency, near the BZ center,  similar to the high frequency portion of the QSL response.

\section{Materials overview and unusual magnetism}
\label{sec:materials}

In this section, we briefly discuss some of the relevant materials.\footnote{We can point the reader to a few recent related reviews 
 \cite{Lee2008end,Balents2010spin,WitczakKrempa2014correlatedAnnualReview,Rau2016spinAnnualReview,Zhou2016quantum,Savary2017quantum,Trebst2017kitaev}.}
Note that the various exchanges, necessarily generated by the geometry and spin orbit coupling, complicate the interpretation of a famous standard measure of proximate QSLs\cite{Obradors1988magnetic,Ramirez1994strongly}, namely the so-called ``frustration parameter''. It is defined as the ratio of the Curie-Weiss temperature $T_{CW}$ to the magnetic ordering temperature $T_N$. However since $T_{CW}$ is related to the average of the magnetic exchanges across all bonds, the various bond-dependent exchanges, which may appear with differing signs, can easily cancel each other out to produce an anomalously small or even vanishing  \cite{Reuther2011finite} value for $T_{CW}$. This value can easily underestimate the true value of the frustrated magnetic exchanges.

At low energies, both Na$_2$IrO$_3$ and RuCl$_3$ order into a collinear ordered``zigzag" pattern at wavevector $M$ (edge midpoint of the hexagonal lattice BZ) \cite{Singh2012relevance,Ye2012direct,Liu2011long,Choi2012spin,Sears2015magnetic}. This order is consistent with large Kitaev exchange \cite{Chaloupka2013zigzag} but also with other models such as further-neighbor exchanges \cite{Mazin2012Na2IrO3,Kimchi2011}. In Na$_2$IrO$_3$, an unusual relation between spin and momentum at temperatures above the zigzag ordering transition provides direct evidence for strong Kitaev exchange \cite{Chun2015direct}.

The three structural polytypes $\alpha,\beta,\gamma$-Li$_2$IrO$_3$ all show \cite{Biffin2014unconventional,Biffin2014noncoplanar,Williams2016incommensurate} an extremely unusual magnetic order which appears to be a unique signature of the Kitaev exchange. This magnetic order is depicted in Figure~\ref{fig:KitaevStructureFactor}(c) in its basic mode common to all three polytypes; the materials differ mainly by various additional patterns of tilts of the spin out of the $x,y$ plane. 
The ordering is an incommensurate order at wavevectors near $0.57$ consisting of spin spirals; however, here the spirals on the $A$ and $B$ sublattices of the crystals have \textit{opposite} senses of rotation. The counterrotating spiral order cannot be stabilized by a Hamiltonian based on nearest-neighbor Heisenberg exchange, since the expectation value of those correlations vanishes due to the counter-rotation of adjacent sites. In particular, for the counter-rotating mode,
\begin{equation}
\sum_{\langle i,j\rangle} \langle
\vec{S}_i \cdot \vec{S}_j \rangle_{\textrm{counterrotating-spiral}} = 0
\end{equation}
where $\langle i,j\rangle$ denotes nearest-neighbors. 
Instead, the nearest-neighbor spin correlations are of a Kitaev-like form \cite{Kimchi2015unified}. This can most easily be seen from Figure~\ref{fig:KitaevStructureFactor}(c) by tilting your head 45 degrees, so that the zigzag chain (a structural feature common to all the honeycomb-type lattices) appears as a staircase, and the $x,y$ spin axes shown are horizontal and vertical respectively. Then it becomes evident that  $x$-bonds ($y$-bonds) have aligned $S^x$  ($S^y)$ but anti-aligned $S^y$ ($S^x)$. 
   Indeed variants of this order have been shown to arise from models with strong ferromagnetic Kitaev exchange and smaller additional antiferromagnetic Heisenberg exchange \cite{Lee2015theory,Kimchi2015unified}. Moreover, a lattice-spin transformation (the ``Klein duality") \cite{Khaliullin2002quantum, Chaloupka2010kitaev, Kimchi2014Kitaev, Chaloupka2015hidden}, which maps Heisenberg models to models with strong Kitaev exchange, was shown \cite{Kimchi2016spin} to transform the usual Heisenberg co-rotating spiral into the counter-rotating spiral, demonstrating it has a parent Kitaev-based model.

A number of experiments have measured dynamical features that appear reminiscent of dynamics seen in the Kitaev model. This has been discussed most prominently in the context of $\alpha$-RuCl$_3$ \cite{Plumb2014rucl3}. Raman scattering observed a broad polarization independent magnetic continuum \cite{Sandilands2015scattering,Sandilands2016spin} which would imply Kitaev coupling $J_K\approx 8$meV from comparison to predictions of the pure Kitaev model \cite{Knolle2014raman}. The continuum persists to high temperatures of the order of $J_K$ and the integrated response, with background subtracted, appears to follow the simple form $\left[1-f(T) \right]^2$ with the Fermi function $f(T)$. This has been interpreted as a signature  of spin fractionalization into fermionic degrees of freedom \cite{Nasu2016fermionic}. Similar behavior has also been reported for $\beta$- and $\gamma$-Li$_2$IrO$_3$ \cite{Glamazda2016raman}.

INS results at high frequencies have also been discussed in the context of the Kitaev model dynamics. 
First, results on RuCl$_3$ from powder scattering revealed the presence of a broad high frequency low wavenumber magnetic continuum which is insensitive to cooling through the AFM transition below which only the very low frequency response develops sharp spin-wave-like excitations \cite{Banerjee2016proximate}. Second, measurements on single crystals \cite{Banerjee2016neutron} strikingly revealed a broad star shape like scattering in reciprocal space, again with a central column of scattering around the zone center, whose main part is almost independent of frequency and temperature (again up to $T^{**}\approx J_K \approx 8$ meV).  The high-frequency portion of the phenomenology appears remarkably similar to that of the proximate Kitaev QSL \cite{Knolle2014dynamics,Do2017incarnation} discussed above. 
Other experimental probes, e.g. thermal conductivity \cite{Hirobe2016magnetic,Leahy2016anomalus} and NMR \cite{Yadav2016}, have been interpreted in the same framework. 
The idea that signatures of the proximate QSL survive at intermediate frequency and temperature regimes despite the appearance of residual long range magnetism below $T_N \ll T^{**} \sim J_K$ appears to be similar to the case of quasi-one-dimensional spin chain materials which display dynamical correlations of the fractionalized spinons \cite{Tennant1995measurement,Mourigal2013fractional} despite weak long range order set by the interchain coupling. However, such a generalization should be taken with great care because 1D fractionalization  is qualitatively different  from  $D \geq 2$ fractionalization \cite{Wen2002quantum, Senthil2000Z2gauge}. In 1D there is no confinement-deconfinement transition: spinons are a generic feature of 1D systems in contrast to fractionalization in $D \geq 2$ which involves deconfinement and ``topological order'' \cite{Wen2002quantum}.

Alternative interpretations of these measurements, that do not invoke the spin liquid variables, have also been discussed. As mentioned above, spin waves are sufficient for reproducing the frequency-wavevector location of the scattering, with the broadness of the feature requiring magnons to break down at these high frequencies. 
The good agreement of a recent ED study with the INS results discussed above \cite{Winter2017breakdown} was interpreted in terms of such a magnon breakdown picture for the zigzag order seen in RuCl$_3$.
Whether the natural quasiparticle description of the signal is best described in terms of Majorana fermions or in terms of multi-magnon excitations is a matter of debate in the literature \cite{Winter2017breakdown,Banerjee2016proximate}. 
Nevertheless, since such a signal is not typically seen in most magnetic systems, its presence can be associated with the unusual correlations from the Kitaev exchange. 
Overall, there is growing and solid evidence across the recent literature that these materials must be described by Hamiltonians that include strong Kitaev-type interactions, and thus are in some sense ``proximate'' to the Kitaev QSL phase.

\section{Future directions}
\label{sec:future}

The two most prominent questions asked in this field are: \newline
{\it How can we drive the materials into a QSL regime or otherwise expose physics related to fractionalization?}   \newline
{\it How can we design experiments that show an unambiguous signature unique to a QSL phase?}\newline

Different avenues have been proposed for driving the systems out of the ordered states and into a QSL phase. The preliminary measurements all rely on observations of the disappearance of the magnetic order, under application of pressure \cite{Takayama2015hyperhoneycomb,Breznay2017resonant}, chemical substitution \cite{unpublished}, or by applying external magnetic fields  \cite{Kubota2015successive,Johnson2015monoclinic,Majumdar2015anisotropic,Baek2017observation,Modic2016robust,Ruiz2017field}. 
It is currently still unclear why the magnetic ordering disappears, but one possible explanation could be that the (chemical) pressure distorts the octahedral structure. The latter may be more advantageous for large Kitaev interactions \cite{Winter2016challenges}.
It may be fruitful to look for other materials as well  -- the metal-organic-frameworks \cite{Yamada2016designing} suggest a promising avenue in this direction, especially since they can realize different lattice structures, and thus different Kitaev QSLs, than the Iridates and RuCl$_3$ \cite{Ohrstrom20043Dnets}.

A theoretical quantification of how much fine tuning would be required to reach the QSL ground state is generally unknown. 
Exact diagonalization \cite{Chaloupka2010kitaev} as well as density matrix renormalization group  studies in two dimensions \cite{Jiang2011possible} find that the QSL phases are stable to adding Heisenberg exchange of the order of a few percent. However, for generic interactions it may in fact be less \cite{Rousochatzakis2015phase}.
A tensor network study in effectively infinite dimensions \cite{Kimchi2014three} (on the boundary-less Bethe lattice) found that the gapped anisotropic phase of the QSL at least is stable to Heisenberg exchange of only much less than a percent perturbation. This is, however, not necessarily too discouraging, as the gapped Kitaev QSLs are generically much less stable than the gapless ones, because of their substantially smaller flux gap \cite{Pachos2007wavefunction}. The stability of the various gapless 3D Kitaev spin liquids to Heisenberg (or other) interactions is  currently not known. 

Several experimental results discussed earlier  show properties that can be interpreted as stemming from the spin fractionalization to Majorana fermions, even though this interpretation is still under debate. 
Doping mobile charges into the system is also thought to expose physics of fractionalization. 
Doped mobile holes interacting via the QSL background  can induce unconventional superconductivity \cite{You2012doping,Hyart2012competition,Okamoto2013global,Halasz2014doping}.  Moreover doping charges into a magnetic phase that is proximate to a QSL phase may uncover the QSL variables and turn the fractionalization physics into the correct description at finite doping  \cite{You2012doping}. 

On the theory side, it is important to further develop the phenomenology both for QSL phases as well as for the various magnetic phases with strong spin orbit coupling. 
This would enable a distinction between unusual signatures of ``proximate'' QSL behavior and of more conventional ordered magnetic phases. 
Experiments that can tune through parameter space, e.g.\ via pressure or even strain, may be the most promising for finding a QSL ground state. 
The recent surge in experimental efforts related to the physics of the Kitaev QSL, including synthesis of new materials, raises the hope that the near future will see many advances in the search for these elusive quantum states, and hopefully even the first unambiguously clear determination of a QSL material. 
 
\vspace{.2cm}

{\bf ACKNOWELDGMENTS:}  
We thank J. Analytis, W. Brenig, K. Burch, C. Castelnovo, K.-Y. Choi R. Coldea, P. Gegenwart, G. Jackeli, G. Khalilullin, Y.-B. Kim, P. Lemmens, Y. Motome, J. Pachos, F. Pollmann, S. Trebst, A. Vishwanath, and M. Vojta for insightful discussion. 
I.K. thanks J. Analytis, N. Breznay, R. Coldea, A. Frano, J. Hinton, G. Jackeli, Sundae Ji, R. Jonnson, G. Khalilullin, K. Modic, J. Orenstein, J.-H. Park, S. Patankar, A. Ruiz, L. Sandilands, T. Smidt, A. Vishwanath, Y-Z You, and other group members for related collaborations and many discussion. 
J.K. is indebted to R. Moessner, D. L. Kovrizhin and J. T. Chalker who have shaped his understanding of the field.
J.K. would like to thank A. Smith, J. Nasu, Y. Motome, B. Perreault, F. J. Burnell, N. B. Perkins,  S. Bhattacharjee, S. Rachel, G. W. Chern, I. Rousochatzakis, S. Kourtis, as well as, S. Nagler, A. Banerjee and A. Tennant for collaborations related to this work. 
M.H. acknowledges partial support through the Emmy-Noether program and CRC 1238 of the DFG. 
I.K. acknowledges support from the MIT Pappalardo Fellowship program. 
J.K. is supported by the Marie Curie Programme under EC Grant agreements No.703697.

\bibliographystyle{unsrt}
\bibliography{references}

\end{document}